\documentclass[12pt]{article}

\usepackage{graphics}
\usepackage{amsmath}
\def\ni{\noindent}
\def\ph{{\phantom{...}}}
\def\={\phantom{..} = \phantom{..}}

\def\+{\phantom{..} + \phantom{..}}
\def\>{\phantom{..} > \phantom{..}}
\def\<{\phantom{..} < \phantom{..}}
\def\-{\phantom{..} - \phantom{..}}

\def\bq{\begin{quote}}
\def\eq{\end{quote}}
\def\be{\begin{equation}}
\def\ee{\end{equation}}
\def\bar{\begin{eqnarray}}
\def\ear{\end{eqnarray}}
\def\no{\nonumber}

\def\Sch{Schr{\"o}dinger}

\def\Schist{Schr{\"o}dingerist}
\def\Schists{Schr{\"o}dingerists}

\def\Copist{Copenhagenist}
\def\Copists{Copenhagenists}

\def\vN{von Neumann}

\def\wf{wavefunction}
\def\wfs{wavefunctions}

\def\MP{Measurement Problem}
\def\qm{quantum mechanics}

\def\enis{energy inequalities}

\def\prodiN{\prod_{i=1}^N}

\def\sijN{\sum_{i,j=1}^N}

\def\srmo{{\sqrt{ - 1}}}
\def\half{{\frac{1}{2}}}
\def\stwfs{space-time wavefunctions}
\def\sts{space-times}
\def\st{space-time}
\def\bt{bitensor}
\def\coor{co{\"o}rdinate}

\def\emt{energy-momentum tensor}
\def\HE{H\&E}
\def\La{Lagrangian} 
\def\cL{{\cal L}}
 \def\btd{{\bigtriangledown}}

\title{\bf On Non-Linear Quantum Mechanics,\\[1in] Space-Time Wavefunctions, \\[1in] 
and Compatibility with General Relativity\\[2in]}

\author{W. David Wick\footnote{email: wdavid.wick@gmail.com}}

\begin{document}
\maketitle
\pagebreak

\section*{Abstract}
In previous papers I expounded non-linear \Schist\ \qm\ as a solution of the \MP. Here 
I show that NLQM is compatible with Einstein's theory of General Relativity.
The extension to curved \sts\ presumes adoption of ``\stwfs" 
(sometimes called ``multi-time \wfs") 
and some additional algebraic structure: a ``\bt" supplementing Einstein's metric tensor.
This kind of matter may violate the Strong Energy Condition even without a mass term, possibly
with implications for the formation of singularities within Black Holes.
 
\section{Introduction}

Reconciling Quantum Mechanics (QM) and Einstein's General Relativity (GR)
has vexed physicists for nearly a century. 
Radical or hegemonic solutions have been proposed (e.g., abandoning Einstein's geometrical axioms 
and deriving gravity from quantum principles of some variety) or at least desired 
(e.g., Einstein's late-life search for a classical field theory combining 
gravity and electromagnetism 
that he presumably hoped would spawn quantum phenomena as a side effect). 
Here I pursue the more modest goal
of demonstrating the compatibility of \Schist\ nonlinear \qm\  
(acronym, NLQM; developed in a series of papers beginning in 2017, see \cite{wdw}),
 with General Relativity  
by contructing one theory of the former kind on a general class of curved \sts. 

Recall Einstein's lament about the structure of his theory: 
as like a house with two wings, one built of the finest marble 
and the other of the cheapest wood. Einstein's gravitational equation
reads:

\be
G_{\mu,\nu} \= (\hbox{constant}) \,T_{\mu,\nu}.\label{eeqn}
\ee

\ni On the left side we find Einstein's curvature tensor and on the right side 
the energy-momentum (also known as stress-energy) tensor of matter. 
But after the 1920's matter was presumed quantum in some sense 
and Einstein never found a quantum theory of matter (which, presuming by ``finest marble" he meant 
the left side and by ``cheapest wood" the right side, explains his discomfiture). 

I will define ``compatibility" to mean a quantum theory on curved \st\ 
that is (a) invariant under general \coor\ transformations, and (b) defines an \emt\
that is conserved and so can be placed on the right side of 
Einstein's equation (\ref{eeqn}). Both the \emt\ and the dynamical equations for the quantum \wf\
will involve the metric tensor $g_{\mu,\nu}$. Thus, as Einstein supposed, 
gravity will influence matter, which in turn will influence gravity. 

This extension of NLQM requires two developments, neither original here. 
Throughout the paper I will write ``events" (points of \st) as $x$, $y$ etc., confounding somewhat 
with the description of an event as a four-tuple, e.g., $x = (x^0,x^1,x^2,x^3)$, 
in some local \coor\ 
system. First development: the replacement of single-time \wfs\ with functions of 
$N$ space-time events, i.e., of form $\psi(x_1,x_2,...,x_N)$ where each variable denotes 
a four-tuple. This step is natural, as insisting on a foliation of \st\ by space-like 
three-manifolds which can represent ``all of space at a given time" violates the spirit of GR. 
I will refer to them as ``space-time" rather than as ``multi-time" \wfs, as the latter 
might be confused with some higher-dimensional redefinition of time. 
These \wfs\ represent a strict generalization of the usual class, 
since if you possess that foliation you can obtain a familiar single-time 
\wf\ merely by restriction. 
Space-time \wfs\ appeared briefly in a paper of Dirac and colleagues in 1932, \cite{dfp}, 
and have been discussed more recently by Lineart, Petrat, and Tumulka in 2017, \cite{lpt},
 who advertised their 
utility in particle theory (not relevant here; these authors also did not entertain 
the full expansion of the class, which may produce new phenomena 
such as ``time-cats", see Discussion section).

The second development is a ``\bt", which I will write as $h_{\mu,\nu}(x,y)$ 
and which, in contrast to the metric tensor 
$g_{\mu,\nu}(x)$, is a function of two space-time points. The necessity of postulating a bitensor
essentially follows from the basic distinction between GR and QM: the former is purely local
(equating tensors, which act on local objects, tangent vectors at a point), 
while the latter of course is nonlocal. I provide more concrete motivation in the next section,
and technical and historical material in an Appendix. 

For simplicity of exposition, I restrict attention in this paper to scalar complex-valued \wfs.
Increasing the number of components (e.g., to incorporate ``spin") appears to be 
straightforward, but can await a later work. I also set all physical constants (including $\hbar$,
$G$, and $c$) equal to one; the reader will easily restore them to their proper places if desired.

I will follow the Hilbert invariant-Lagrangian program rather than the Hamiltonian approach 
of the Measurement Problem papers, as it automatically 
incorporates Einstein's curvature tensor and gravitational equation.
 
\section{The Bitensor\label{btsection}}

The motivation for introducing the \bt\ derives from contemplating an expression like:

\be
\sijN\,<\psi|\,g^{\mu,\nu}\,\left\{\,P_{i,\mu} \- <\psi|\,P_{i,\mu}\,|\psi>\,\right\}
\,\left\{\,P_{j,\nu} \- <\psi|\,P_{j,\nu}\,|\psi>\,\right\}\,|\psi>,\label{MPexp}
\ee

\ni where

\bar
\no P_{i,\mu} &\=& \srmo\,\partial_{i,\mu}\\
\no &\=& \srmo\,\partial/\partial\,x_{i}^{\mu},
\ear

\ni which arises in the nonlinear quantum theory proposed in the first of a series of papers,
\cite{wdw}. Developing this expression will generate terms like:

\be
<\psi|\,g^{\mu,\nu}\,P_{i,\mu}\, P_{j,\nu}\,|\psi>,
\ee

\ni which does not have an obvious meaning on curved manifolds because two (co-) vectors 
referenced to different \st\ points are involved; i.e., do we put $g^{\mu,\nu}(x_i)$ 
or $g^{\mu,\nu}(x_j)$? Also, of course, what meaning to give to $<\,\psi\,|\,\,|\,\psi\,>$,
a question which provides additional motivation for switching to \st\ \wfs.

Thus one problem to be overcome is how to compare vectors 
(or covectors) attached to different points,
say x and y, on a curved manifold. On a flat manifold the answer is obvious: 
parallel-transport the vector at x over to y and compare using the constant metric tensor 
(or vice-versa, which gives the same thing). Of course, the very definition of curvature is that 
it renders such a comparison curve-dependent. 

Hence the motivation for the \bt. We require natural properties suggested by the terminology:
first, that it reduces to the metric tensor when the arguments coincide:

\be
h^{\mu,\nu}(x,x) = g^{\mu,\nu}(x) \label{bt1}.
\ee

\ni Second, that it be tensorial in each variable and jointly: if $'$ denotes a ``primed" \coor\ system,

\bar
\no h^{'\mu,\nu}(x',y) &\=& \frac{\partial\,x'_{\mu}}{\partial\,x_{\alpha}}\,h^{\alpha,\nu}(x,y);\\
\no h^{'\mu,\nu}(x,y') &\=& \frac{\partial\,y'_{\nu}}{\partial\,y_{\alpha}}\,h^{\mu,\alpha}(x,y);\\
 h^{'\mu,\nu}(x',y') &\=& \frac{\partial\,x'_{\mu}}{\partial\,x_{\alpha}}\,
  \frac{\partial\,y'_{\nu}}{\partial\,y_{\beta}}\,h^{\alpha,\beta}(x,y);\label{bt2}
\ear

\ni (usual summation convention on Greek letters applies). 

Third, the natural symmetry properties hold:

\be
h^{\mu,\nu}(x,y) = h^{\nu,\mu}(y,x) \label{bt3}.
\ee

\ni I have chosen to relate contravariant tensors 
because they will appear in the quantum Lagrangian (next section). Covariant tensors  
 are available if required by the usual lowering of indices using the 
local metric for each argument; e.g.,

\be h_{\mu,\nu}(x,y) \= g_{\mu,\alpha}(x)\,g_{\nu,\beta}(y)\,h^{\alpha,\beta}(x,y).
\ee

Given a four-dimensional manifold and a pseudo-Reimannian metric tensor, does such a \bt\ exist?
The answer is, always; existence can be derived from 
the embedding theorems into higher-dimensional flat (pseudo-Euclidean) spaces proven in the 1970s. 
However, the techniques involved
vary from author to author; moreover, we do not need 
the full strength of these difficult theorems---we only need, so to speak, 
the derivatives of the mappings, 
not the mappings themselves. 
Nevertheless, to establish a rigid, predictive theory we should desire a definite formulation. 
That can be provided for ``generic" space-times 
satisfying a geodesic-connectivity condition and
by relaxing the requirement that 
$h$ be defined for every pair $(x,y)$, replacing ``every" by ``almost every". 
All of this is discussed in the Appendix.

\section{The Lagrangian\label{lagrangesect}}

I will assume symmetry of the wavefunction under permutation of arguments, 
which makes for more compact mathematical expressions but is not essential.
I also omit from the Lagrangian mass terms, interaction terms, and external potentials, 
in order to focus on what is new (nonlinear and nonlocal terms).
 
We will require some definitions of symbolic expressions. First, for the volume element 
on the manifold $M$ (four-dimensional without boundary; \st) and and its product on $M\times M$:

\bar
\no dv(x) &\=& d^4x\,\sqrt{-g(x)};\\
 dv_2(x,y) &\=& d^4x\,\sqrt{-g(x)}\,d^4y\,\sqrt{-g(y)};
\ear

\ni where $g$ denotes the determinant of the covariant metric tensor $g_{\mu,\nu}$.
Also, for integrating out variables,

\bar
\no <A\psi|B\psi>^{(1)}(x) &\=& \prod_{i=1}^{N-1}\,\int\,dv(x_i)\,A\psi^*(x_1,x_2,..,x_{N-1},x)\,
B\psi(x_1,x_2,..,x_{N-1},x);\\
\no <A\psi|B\psi>^{(2)}(x,y) &\=& \prod_{i=1}^{N-2}\,\int\,dv(x_i)\,
A\psi^*(x_1,x_2,..,x_{N-2},x,y)\,\times\\
&& \ph\ph\ph\ph B\psi(x_1,x_2,..,x_{N-2},x,y);
\ear

\ni where $^*$ denotes complex conjugation and $A$ and $B$ are any real operators, 
say differential operators.

The overall \La\ will 
take the familiar form (a good reference here, which I will cite repeatedly,
is Hawking and Ellis 1973, \cite{hande}, abbreviated below H\&E, see their section 3.3): 

\be
\cL \= \cL_g \+ \cL_{\psi},\label{laeq}
\ee

\ni that is, a sum of a term involving only the metric tensor and a term involving
the metric tensor and the \wf.
The first term is Hilbert's invariant Lagrangian for gravity:

\be
\cL_g \= \frac{1}{2\pi}\,\int\,dv(x)\,R(x),
\ee

\ni where $R(x)$ is the Ricci scalar. For the second term in (\ref{laeq}) we will entertain
combinations of terms of three types, which we label with a prefactor of $a$, $b$, or $c$. 
First, a term rationalizing an $i\neq j$ term from (\ref{MPexp}):

\be
a\,\int\,\int\,dv_2(x,y)\,h^{\mu,\nu}(x,y)\,<\,\partial_{x,\mu}\psi\,|\,
\partial_{y,\nu}\psi\,>^{(2)}(x,y);\label{Lagrangiana}
\ee

Second, a term from that same expression but quartic in $\psi$:

\be
b\,\int\,\int\,dv_2(x,y)\,h^{\mu,\nu}(x,y)\,<\,\partial_{x,\mu}\psi\,|\,\psi\,>^{(1)}\,
<\,\psi\,|\,\partial_{y,\nu}\,\psi\,>^{(1)};\label{Lagrangianb}
\ee

Third, a term coming from that expression for $i=j$:

\be
c\,\int\,dv(x)\,g^{\mu,\nu}(x)\,<\,\partial_{x,\mu}\,\psi\,|\,
\partial_{x,\nu}\,\psi\,>^{(1)}.\label{Lagrangianc}
\ee

The form of these expressions was chosen so that, in the Minkowski (flat-space) limit,
they permit a reassembly of expression (\ref{MPexp}), see section \ref{flatsect}.

The dynamical equations for the \wf\ derives from from varying the Lagrangian
 with respect to $\psi$,
while the \emt\ and Einstein's equation derives from varying it with respect to the metric.

\section{The Dynamical Equations for the Wavefunction\label{dynsect}}

We vary the Lagrangian with respect to $\psi$ while keeping $g_{\mu,\nu}$ constant. 
Accordingly let $\psi(\cdot,u)$ denote a one parameter
family of \wfs\ with $\psi(\cdot,0) = \psi(\cdot)$. 
We will need an expression for the derivative of the volume element, which involves 
$g = \hbox{det}(g_{\mu,\nu})$,  with respect to
some \coor, which we can derive following \HE, p. 166:

\bar
\no \partial_{x,\mu}\,\sqrt{-g} &\=& - \half\,(-g)^{-1/2}\,g\,
\frac{\partial\,g}{\partial\,g_{\alpha,\beta}}\,\partial_{x,\mu}\,g_{\alpha,\beta}\\ 
&\=& \half\,\sqrt{-g}\,g^{\alpha,\beta}\,\partial_{x,\mu}\,g_{\alpha,\beta}.\label{heeq}
\ear

We write $\Delta \psi$ for $d/du\,\psi$. The goal is an
expression:

\be
\frac{d\,\cL_{\psi}}{du} \= \prodiN\,\int\,dv(x_i)\,
\left\{\,\Delta\,\psi^*\,D\,\psi\, \+
\Delta\,\psi\,D\,\psi^*\,\right\};
\ee

\ni which we require to be zero for arbitrary $\Delta\,\psi$.
For each term in the Lagrangian we integrate-by-parts whenever 
we see derivatives on $\Delta\,\psi^*$, obtaining a contribution
labelled 
$D_a\psi$, etc, which sum to $D\,\psi$. For the term with prefactor `a', omitting an overall minus sign:  

\bar
\no  D_a\,\psi &\=&
 \half\, h^{\mu,\nu}(x,y)\,\,g^{\alpha,\beta}(x)\,
\partial_{x,\mu}\,g_{\alpha,\beta}(x)\,
\partial_{y,\nu}\,\psi \+\\
\no && \partial_{x,\mu} h^{\mu,\nu}(x,y)\,\partial_{y,\nu}\psi \+\\
&& \,h^{\mu,\nu}(x,y)\,\partial_{x,\mu}\,\partial_{y,\nu}\,\psi.\label{dyneqa}
\ear

Thus the `$a$' term in the Lagrangian
yields a linear, Klein-Gordon type equation with first-order terms. 
(The equation singles out $x,y$,
standing in for $x_{N-1},x_N$; but, again using symmetries, we can rewrite it as an average
of $N(N-1)$ terms with derivatives with respect to $x_i,x_j$ for $i \neq j$.) 

Equation 
(\ref{dyneqa}) can be written more elegantly by introducing the covariant derivative
operator connected to the metric, defined on, e.g., a scalar and a tensor by:

\bar
\no \btd_{\mu,x}\,\psi(x) &\=& \partial_{\mu,x}\,\psi(x);\\
\no \btd_{\mu,x}\,K^{\nu,\alpha}(x) &\=& \partial_{\mu,x}\,K^{\nu,\alpha} +
\Gamma^{\nu}_{\beta,\mu}\,K^{\beta,\alpha} +
\Gamma^{\alpha}_{\beta,\mu}\,K^{\beta,\nu},\\
&& 
\ear

\ni where the $\Gamma^{\alpha}_{\beta,\mu}$ are the connection symbols. We can exploit
also a formula for, e.g., the divergence of a vector:

\be
\btd_{\mu,x}\,V^{\mu}(x) \= \left[\,\sqrt{-g(x)}\,\right]^{-1}\,\partial_{\mu,x}\,\left[\,\sqrt{-g(x)}\,
V^{\mu}(x)\,\right].
\ee

Thus we can rewrite (\ref{dyneqa}) as:

\bar
\no  D_a\,\psi &\=&
\btd_{x,\mu} h^{\mu,\nu}(x,y)\,\btd_{y,\nu}\psi \+\\
\no && h^{\mu,\nu}(x,y)\,\btd_{x,\mu}\,\btd_{y,\nu}\,\psi.\\
&&
\ear

Similarly, for '$b$' terms 
we find:

\bar
\no  D_b\,\psi &\=&
 \int\,dv(y)\, \btd_{x,\mu}\,h^{\mu,\nu}(x,y)\,<\,\psi\,|\,\partial_{y,\mu}\,
\psi\,>^{(1)}\,\psi  \\
\no && \+  \int\,dv(y)\, \,h^{\mu,\nu}(x,y)\,<\,\psi\,|\,\partial_{y,\mu}\,
\psi\,>^{(1)}\,\partial_{x,\mu}\,\psi \\
\no && \-  \int\,dv(x)\, \,h^{\mu,\nu}(x,y)\,<\,\partial_{x,\mu}\,\psi\,|\,
\psi\,>^{(1)}\,\partial_{y,\mu}\,\psi \\
&& \label{dyneqb}
\ear

Thus including `$b$' terms renders the dynamical equations nonlinear (cubic in $\psi$) 
and containing integrations.

Finally, from the terms containing `$c$':

\bar
\no  D_c\,\psi &\=& 
\btd_{x,\mu}\,g^{\mu,\nu}(x)\,\btd_{x,\nu}\,\psi \+ \\
\no && g^{\mu,\nu}(x)\,\btd_{x,\mu}\,\btd_{x,\nu}\,\psi\\
\no &\=& g^{\mu,\nu}(x)\,\btd_{x,\mu}\,\btd_{x,\nu}\,\psi.\\
&&\label{dyneqc}
\ear

\ni That the first term after the equality
 vanishes follows from the compatibility between the metric and the derivative operator, 
\HE, equation (2.25):

\be
\btd_{x,\alpha}\,g^{\mu,\nu}(x) \= 0.
\ee

With all three kind of terms, the full dynamical equations become:

\be 
D\,\psi \= \left\{\,D_a + D_b + D_c\,\right\}\,\psi \= 0,
\ee

\ni because the variation must vanish for arbitrary $\Delta\,\psi$.
 
\section{The Energy-Momentum Tensor, Conservation Laws, and Einstein's Equation\label{emtsect}}

We now vary the (inverse) metric tensor keeping the \wf\ fixed.
Accordingly let $g^{\mu,\nu}(x;u)$ denote a one-parameter family of metrics with 
$g^{\mu,\nu}(x;0) = g^{\mu,\nu}(x)$, and again we use $\Delta$ for $d/du|_0$. 
We will need a formula for the variation of the bitensor:

\be
\Delta\,h^{\mu,\nu}(x,y) \= \int\,dv(z)\,\frac{\delta\,h^{\mu,\nu}(x,y)}{\delta\,g^{\alpha,\beta}(z)}\,
\Delta\,g^{\alpha,\beta}(z).\label{star}
\ee

It is possible that the integrand here is singular; that is, contains Dirac delta-functions, see
Appendix.
We find for terms involving `$a$':

\bar
\no \frac{d\,\cL_{\psi}}{du} &\=& a\,\prod_{i=1}^{N-2}\,\int \,dv(x_i)\,\int\, \int \,dv_2(x,y)
\partial_{x,\mu}\,\psi^*\,\partial_{y,\nu}\,\psi\, 
\left\{\,\Delta\,h^{\mu,\nu}(x,y) \+\right.\\
\no&& h^{\mu,\nu}(x,y)\,
\half\,\left[\,\sum_{i=1}^{N-2}\,
g_{\alpha,\beta}(x_i)\, \Delta\,g^{\alpha,\beta}(x_i)\right.\\
\no &&\left.\left. \+ g_{\alpha,\beta}(x)\, \Delta\,g^{\alpha,\beta}(x) \+ 
g_{\alpha,\beta}(y)\, \Delta\,g^{\alpha,\beta}(y)\, 
\right]\,\right\}\\
&& \label{emta}
\ear

\ni which we must rewrite in the form,
using (\ref{heeq}) and (\ref{star}):

\be
\half\,\int\,dv(z)\,T_{\alpha,\beta}(z)\,\Delta\,g^{\alpha,\beta}(z).
\ee

Thus for the `$a$' terms we can write

\bar
\no T_{\alpha,\beta}(z)_a &\=& 2\,a\int\,\int\,dv_2(x,y)
\,<\partial_{x,\mu}\,\psi\,|
\,\partial_{y,\nu}\,\psi>^{(2)}  
\,\frac{\delta\, h^{\mu,\nu}(x,y)}{\delta\,g^{\alpha,\beta}(z)} \-\\
\no && \,g_{\alpha,\beta}(z)\,I_a(z);\\
&&\label{emta1}\\
\no I_a(z) &\=& a\,\int\,\int\,dv_2(x,y)\,h^{\mu,\nu}(x,y)\,
\left\{\,\sum_{i=1}^{N-2}\,\prod_{j\neq i}\,\int\,dv(x_j)
\left[\,\partial_{x,\mu}\,\psi^*\,\partial_{y,\nu}\,\psi\,\right]_{x_i = z}\right\}\+\\
\no && a\,\int\,dv(y)\,h^{\mu,\nu}(z,y)\left[\,<\,\partial_{x,\mu}\,\psi\,|\,
\partial_{y,\nu}\,\psi\,>^{(2)}\,\right]_{x=z} \+\\
\no && a\,\int\,dv(x)\,h^{\mu,\nu}(x,z)\left[\,<\,\partial_{x,\mu}\,\psi\,|\,
\partial_{y,\nu}\,\psi\,>^{(2)}\,\right]_{y=z}\\
&&\label{emta2}
\ear

For the `$b$' terms we get similarily:

\bar
\no T_{\alpha,\beta}(z)_b &\=& 2\,b\int\,\int\,dv_2(x,y)
\,<\partial_{x,\mu}\,\psi\,|\,\psi\,>^{(1)}
\,<\,\psi\,|\,\partial_{y,\nu}\,\psi\,>^{(1)}  
\,\frac{\delta\, h^{\mu,\nu}(x,y)}{\delta\,g^{\alpha,\beta}(z)} \-\\
\no && g_{\alpha,\beta}(z)\,I_b(z);\\
&&\label{emtb1}\\
\no I_b(z) &\=& b\,\int\,\int\,dv_2(x,y)\,h^{\mu,\nu}(x,y)\,\times\\
\no && \sum_{i=1}^{N-1}\,\prod_{j\neq i}\,\int\,dv(x_j)\,\left\{\,
\left[\,\partial_{x,\mu}\,\psi^*\,\psi\,
\right]_{x_i = z}\,
<\,\psi\,|\,\partial_{y,\nu}\,\psi\,>^{(1)} \+\right.\\
\no && \left.
\left[\,\psi^*\,\partial_{y,\mu}\,\psi\,
\right]_{x_i = z}\,
<\partial_{x,\nu}\,\psi\,|\,\psi\,>^{(1)}\,\right\} \+\\
\no && b\,\left\{\, \int\,dv(y)\,h^{\mu,\nu}(z,y)
\left[\,<\,\partial_{x,\mu}\,\psi\,|\,\psi\,>^{(1)}\,\right]_{x=z} \,
<\,\psi\,|\,\partial_{y,\nu}\,\psi\,>^{(1)}\+\right.\\
\no && \left. \int\,dv(x)\,h^{\mu,\nu}(x,z)\,<\,\partial_{x,\mu}\,\psi\,|\,\psi\,>^{(1)}\,\left[
<\,\psi\,|\,\partial_{y,\nu}\,\psi\,>^{(1)}\,\right]_{y=z}\,\right\}\\
&&\label{emtb2}
\ear

For the `$c$' terms:

\bar
\no T_{\alpha,\beta}(z)_c &\=& c\,\left\{
<\partial_{z,\alpha}\,\psi\,|
\,\partial_{z,\beta}\,\psi>^{(1)} \-
\half\,g_{\alpha,\beta}(z)\,g^{\mu,\nu}(z)\, 
<\partial_{z,\mu}\,\psi\,|
\,\partial_{z,\nu}\,\psi>^{(1)}\,\right\}\\ 
\no && \-\,g_{\alpha,\beta}(z)\,I_c(z);\\
\no I_c(z) &\=& 
 c\,\int\,dv(x)\,g^{\mu,\nu}(x)\,\left(\,\sum_{i=1}^{N-1}\,\prod_{j\neq i}\,
\int\,dv(x_j)\,\left[\,\partial_{x,\mu}\,\psi^*\,
\partial_{x,\nu}\,\psi\,\right]_{x_i = z}\, \right).\\
&&\label{emtc2}
\ear

(Note: the minus signs in front of the terms involving $I_{.}(z)$ arise because in this section
we differentiate with respect to $g^{\mu,\nu}$ rather than its inverse.)

If all three types of terms are in the Lagrangian, the full \emt\ is a sum of the expressions above:

\be 
T_{\alpha,\beta}(z) \= T_{\alpha,\beta}(z)_a + T_{\alpha,\beta}(z)_b + T_{\alpha,\beta}(z)_c.
\ee

We will see in later sections that a considerable simplification in the terms of the Lagrangian
accrues from consideration of the dynamical equations and from plausible 
assumptions about the bivector (introduced in the Appendix).

From this set-up the conservation of energy and momentum now follow automatically,
as in \HE, p. 67, by considering a variation induced by a local diffeomorphism of M,
say induced by a vector field $X$ with compact support. Writing 

\be
\cL_{\psi} \= \prodiN\,\int\,dv(x_i)\,L_{\psi},
\ee

We find

\bar
\no \delta\,\cL_{\psi} &\=& \prodiN\,\int\,dv(x_i)\,\frac{\delta\,L_{\psi}}{\delta\,
\psi(x_1,...,x_N)}\,\Delta\psi(x_1,...,x_N) \+\\
&& \half\,\int\,dv(z)\,T_{\alpha,\beta}(z)\,\Delta\,g^{\alpha,\beta}(z).
\ear

The first term vanishes by the dynamical equations and we can write in the second term:

\be
 \Delta\,g^{\alpha,\beta} \= 2\,\bigtriangledown^{\beta}\,X^{\alpha},
\ee

\ni (see \HE, p. 67). Thus the remaining term can be rewritten as:

\be
\half\,\int\,dv(z)\,\left\{\,\bigtriangledown^{\beta}\,\left(\,T_{\alpha,\beta}(z)\,
X^{\alpha}\right) \- 
\left(\,\bigtriangledown^{\beta}\,T_{\alpha,\beta}\,\right)\,X^{\alpha}\,\right\}.
\ee

By the divergence theorem the first term can be converted to a surface integral 
where $X = 0$ and so vanishes, and since the variation must be zero for arbitrary $X$ we find

\be
\bigtriangledown^{\beta}T_{\alpha,\beta} \= 0,\label{conseq}
\ee

\ni which is our conservation law. 

Einstein's equation follow from varying $g^{\mu,\nu}$ in the full Lagrangian, 
defined in (\ref{laeq}).
Variation of the first term yields: $-G_{\mu,\nu}$, and the second: 
$(\hbox{constant})\,T_{\mu,\nu}$,
thus yielding Einstein's equation (\ref{eeqn}).

\section{The Flat-Space (Minkowski) Limit\label{flatsect}}

In this section I discuss the above scheme in the limit of flat (Minkowski) space. 
Although the concept violates the logic of this paper (if $G_{\mu,\nu} = 0$, by Einstein's equation
also $T_{\mu,\nu} = 0$ which should imply $\psi = 0$), 
it can be regarded as a useful approximation for situations in which
gravity is assumed unimportant for the quantum dynamics. Also, it permits a comparison to the
proposals made in paper I of the \MP\ series.

For the Minkowski limit we put 

\be
h^{\mu,\nu}(x,y) \equiv g^{\mu,\nu}(x) \equiv \eta^{\mu,\nu},
\ee

\ni where $\eta = \hbox{diag}(-1,1,1,1)$ is a constant matrix. Following the sequence of
developments of previous chapters, the Lagrangian has terms of form:

\be
a\,g^{\mu,\nu}\,<\partial_{x,\mu}\,\psi\,|\,\partial_{y,\nu}\,\psi\,>; 
\ee

\be
b\,g^{\mu,\nu}\,<\partial_{x,\mu}\,\psi\,|\,\psi\,>\,<\,\psi\,|\,\partial_{y,\nu}\,\psi\,>; 
\ee

\ni and

\be
c\,g^{\mu,\nu}\,<\partial_{x,\mu}\,\psi\,|\,\partial_{x,\nu}\,\psi\,>. 
\ee

\ni Here, $<\cdot|\cdot>$ stands for integration over everything.

Returning to (\ref{MPexp}), let's write it for this case as:

\be
\sum_{i,j=1}^N\,g^{\mu,\nu}\,\left\{\,<\,P_{i,\mu}\,\psi\,|\,P_{j,\nu}\,\psi\,> \- 
<\,P_{i,\mu}\,\psi\,|\,\psi\,>\,<\,\psi\,|\,P_{j,\nu}\,\psi\,> \,\right\}.
\ee

Exploiting the assumed symmetries of the wavefunction, we can reproduce this expression
by choosing:

\bar
\no c &\=&  N\,\times\,\hbox{constant};\\ 
\no b/c &\=& - N;\\
\no a/c &\=& N - 1.\\
&&
\ear

Next, consider the dynamical equations. Since $g = -1$ is constant, all terms derived from
IBP on the volume element $dv$ are absent. Starting directly from the Lagrangian, 
as opposed to from equations (\ref{dyneqa})--(\ref{dyneqc}), yields: 

\bar
\no D\,\psi &\=& g^{\mu,\nu}\,\left\{\,a\,\btd_{x,\mu}\,\btd_{y,\mu}\,\psi \+\right.\\
\no &&\left. \- b\,<\,\psi\,|\,\btd_{y,\mu}\,\psi\,> \btd_{x,\mu}\,\psi\+ 
 b\,<\,\psi\,|\,\btd_{x,\mu}\,\psi\,> \btd_{y,\mu}\,\psi 
\+\right.\\
\no &&\left. c\, \btd_{x,\mu}\,\btd_{x,\nu}\,\psi\,\right\}\\
\no && = 0.\\
&&
\ear

\ni With the above choices of parameters we can write a more symmetrical version:

\be
D\psi \= \sum_{i,j=1}^N\,g^{\mu,\nu}\,\btd_{x_i,\mu}\,\left\{\,\btd_{x_j,\mu} - 
\gamma_{i,j}\,<\,\psi\,|\,\btd_{x_j,\mu}\,\psi\,>\,\right\}\psi \= 0.
\ee

\ni where $\gamma_{i,j}$ is the unit antisymmetric symbol: 
$\gamma_{i,j} = 1$ if $i > j$, and $\gamma_{i,j} = - \gamma_{j,i}$.

Finally, for the \emt\ only the `$c$' terms survive, yielding:

\bar
\no T_{\alpha,\beta}(z) &\=& c\,\left\{\,
<\partial_{z,\alpha}\,\psi\,|\,\partial_{z,\beta}\,\psi\,>^{(1)} \-\right.\\ 
\no &&\left. \frac{1}{2}\,g_{\alpha,\beta}\,g^{\mu,\nu}\,<\partial_{z,\mu}\,\psi\,|\,
\partial_{z,\nu}\,\psi\,>^{(1)}
\right\},\\
&&
\ear

\ni which the reader will note resembles the usual Lagrangian for the $N = 1$ scalar field,
but with integration over the supplementary variables.

 Re connecting with earlier work: recall that, in classical mechanics, the Lagrangian
is a difference $K - P$, kinetic minus potential energies. In a realistic quantum model
there will presumably be kinetic energy terms involving the mass, 
so perhaps we can regard the contribution
of the terms considered here as representing potential energy. In which case,
conservation laws should apply, limiting the formation of cats, as in paper I of the MP series.

\section{Locality and Causality\label{caussect}}

Although the conservation law is satisfied, 
the commonly-assumed locality condition, {\em viz}., \HE, p. 61 (in our notation):

\begin{quote}
$T^{\mu,\nu}$ vanishes on an open set ${\cal U}$ if and only if 
all the matter fields vanish on ${\cal U}$,
\end{quote}

\ni evidently can be violated in the theory presented here, 
see section \ref{dynsect} and some candidate expressions for the \emt. 

This violation of locality derives solely from the terms
involving derivatives of the bitensor with respect to the metric. 
However, by adopting a construction advertised in the Appendix, all terms 
in $T^{\mu,\nu}$ containing 
$\delta\,h^{\mu,\nu}(x,y)/\delta\,g^{\alpha,\beta}(z)$ vanish. The reason: with the construction,
the derivative will be zero unless $z$ lies on a geodesic connecting $x$ to $y$.
Moreover, for a given $z$, the set of such pairs is of measure zero 
in $M\,\times\,M$ with respect to 
product volume measure.
(Mentally fix $z$ and $x$; then, with the assumptions made in the construction,
the allowable $y$ lie on the union of a finite number of curves,
which has dimension one. Therefore, for given $z$, 
the set of pairs $(x,y)$ cannot have positive measure.)

Does our nonlinear \Schist\ theory satisfy the 
Relativistic Causality Principle (RCP), also known as the 
no-signalling-faster-that-light assumption? The RCP requires that a local perturbation 
in the matter fields should be invisible in measurements 
made in a region space-like separated from the
region of perturbation.
(The concepts of ``space-like" or ``time-like" separated regions 
are well-defined at the manifold-plus-metric level independent of \coor\ systems. 
That is why we can still talk in general relativity
about light-cones, future-and-past of an event, and so forth; 
see \HE, section 6.2.) 
Conventional GR plus local matter fields is known to satisfy
the RCP, see \HE\ section 7.7. But the theory here is nonlocal on the \wf\ side. Might it still 
satisfy the principle?

Let's ask first what should be required of an $N$-argument \wf. To have a definite context,
let's adopt the EPRB set-up. Thus let $A$ and $B$ be regions of \st\ (traditionally 
called ``Alice's and Bob's
laboratories", respectively), which are spacelike separated.
Let $a$ denote
a parameter (say for an external field) controllable by Alice, and $b$ by Bob.
(This is conventional symbolism; $a$ and $b$ are NOT the parameters appearing in the Lagrangian of
section \ref{lagrangesect}.)
For our \wf, the most restrictive definition of the RCP would require that, if any $x_i \in A$,
then $\psi(x_1,...,x_i,...,x_N)$ be unaffected by varying $b$. 

But this definition is too narrow. Given an exact solution of \Sch's equation for EPRB, it will
not be the case that $\psi(x,y)$ is independent of $a$ or $b$; rather, 
both parameters will appear, i.e., the functional form is
$\psi(x,y;a,b)$. The reason that
conventional, linear QM satisfies the RCP in this instance is due to a limitation on allowed
observables. An observation made by Alice is presumed to involve an ``expectation value" 
of an operator containing only ``the spin operator for particle one, which arrived in Alice's
lab"; because the Hamiltonian
is a sum of commuting operators (the two spin operators), for this expectation $b$ drops out.
(In \Copist\ QM the description is articulated by an {\em ad hoc}
random variable having this ``expection value" as mean.) 

Therefore, the issue of satisfying or violating the RCP comes down to allowable observables. 
Nicholas Gisin, Angelo Bassi, GianCarlo Ghirardi and others 
have claimed (see \cite{bandg} for a review) that no nonlinear, deterministic, relativistic
 quantum theory
can satisfy the RCP. But this cannot be a theorem as so stated, 
for there exist known counterexamples.
(For instance, Dirac+Maxwell for one electron, see \cite{barut}; the nonlinearity appears in the
Dirac charge-current placed on the right hand side of Maxwell's equation, 
since it is quadratic rather than linear in $\psi$. Another example: the $N=1$, spin-zero,
\wf\ plus Einstein's metric tensor, 
which is nonlinear because Einstein's equation is, see \HE, p.67, Example 1.) 
The authors of these ``no-go"
theorems assume \vN's measurement axioms: every self-adjoint operator represents an ``observable",
and all observables are of form $<\psi|{\cal A}|\psi>$, 
or $\hbox{tr}\,(\,\rho\,{\cal A}\,)$ if you wish to include mixtures, 
where ${\cal A}$ is such an operator.
\Schists\ will not agree with either requirement (certainly not if posing a nonlinear theory,
for which the Hamiltonian or Lagrangian will contain terms not of \vN's form). Nor must we accept
these author's definition of ``determinism": that if all \vN\ observables agree in a quantum state 
at one time, they will agree at all later times---because we may not agree that observables exist
whose values completely characterise a \wf. (An appropriate measurement philosophy for \Schists\
is ``instrumentalism": an observable only arises if you can describe 
an apparatus, and can model both it and the observed system jointly by a \wf.).

Nevertheless, because the dynamical equations for $\psi$ contain integrals over all of \st, 
satisfying the RCP appears difficult. 
One possible way out appeared earlier in the
\MP\ papers: if the nonlinear terms assemble into something resembling expression (\ref{MPexp}),
their main impact may be on the dispersion of energy-momentum given the wavefunction, 
which is not normally measured in an experiment. (An example in a nonrelativistic setting
appeared in paper I of the \MP\ series: adding the nonlinear, nonlocal terms to the Hamiltonian
there did not affect the motion of the center of mass of an apparatus pointer, 
while, in the presence of the nonlinear terms and at laboratory energies,
the dispersion would be too small to measure.) 
If these terms can be ignored where ordinary measurements and external fields are involved, 
the proof of the RCP may follow from theorems about 
domain-of-dependence for linear hyperbolic systems of PDEs, 
see e.g., \HE\ Chapter 7. 

Alternatively, consider again EPRB and Bob's and Alice's apparatus. If Bob's can
be described by variables $\{\,x_1,x_2,...,x_{N/2}\,\}$ and Alice's by\break
 $\{\,x_{N/2+1},...,x_N\,\}$,
and the wavefunction factors as $\psi_A(x_1,...,x_{N/2})\,\times$ \break 
$\psi_B(x_{N/2+1},...,x_N)$,
a dispersion term would split into a sum (as for variances for sums of independent variables 
in probability theory, which of course isn't a \Schist\ interpretation). Then when
treating of observations involving Bob's lab we could ignore Alice's. Coupling with the 
microsystem (describing the ``pair of particles" for \Copists) 
will generate correlations, but as in
conventional quantum mechanics when Bob measures some local quantity $\phi$, say yielding 
$<\psi|\phi(x_1,...,x_{N/2})|\psi>$, Alice's controllable variable `$a$' may simply drop out.
Thus the theory would, in this case, effectively satisfy the RCP. 

Of course, the correlations
would only become Bell-like after adding a random component to $\psi$, 
which might be {\em ad hoc} or reflect sensitive dependence 
on initial conditions (see papers II and III in the Measurement Problem series).
So another possibility arises: nonlocal effects may exist in the deterministic theory, 
but be obscured by randomness. So in EPRB, 
knowledge of `$a$' and all other relevant conditions might permit a certain (probability one)
 conclusion 
about whether Bob's apparatus-needle center-of-momentum was positive 
(and consequently its center-of-mass moved up). But if what Bob actually observes are random
sequences of up's and down's, it is impossible for him to guess Alice's 
choice from a single observation, possibly not even granted a long series.

In summary: due to the presence of space-time integrals in the dynamical equations, circumstances
could be found which violate the RCP. 
I note that general
relativity also permits such scenarios; for instance, \HE\ show that a certain ``dominant energy
condition" (see next section) must be satisfied to insure 
that matter cannot travel faster than light, see 
p.s 91 {\em et seq.}, especially the remark near the bottom of p. 94. 
Fewster, \cite{fewster}, pointed out that quantum field theory, and even the simple 
$N = 1$ massive scalar field, violate the stronger energy
conditions, and even the weaker requirement 
that the energy density be positive. 
(Fewster and others have argued that such conditions might be true ``on average".)
If the present theory enjoys the DEC, we might argue that faster-than-light 
signaling is impossible (since some transmission of energy must be involved), 
just as we would for GR with ``classical" matter.

\section{Energy Inequalities\label{ineqsect}}

Perhaps the most interesting mathematical results in General Relativity since the founding era 
are the Black Hole and singularity theorems proven in the 1960s
 (fully covered in \HE). Before this period, Black Holes were known only from special solutions
of Einstein's equations with a high degree of symmetry, and it could be argued 
that they were mere mathematical artifacts. The 1960s work established that BH's with attendant 
singularities (where, e.g,
curvature went to infinity), were generic, granted certain reasonable conditions 
imposed on the right side, the \emt.  
These  ``energy inequalities" 
were derived from classical models of matter, say of a normal fluid or gas. But 
matter is supposed to be ultimately quantum, and, as
\HE\ point out (pages 95-6), even for the simplest quantum model, 
the $N = 1$ scalar wavefunction with positive mass, the inequalities may be violated. Which 
raises the question: are the BH and singularity theorems applicable to quantum matter?
In this section I investigate the \enis\ for energy-momentum tensors
 derived from NLQM as I have defined it.

There are three \enis\ that have been employed to prove theorems about space-times:

{\bf The Weak Energy Condition}: for every time-like vector $W^{\alpha}(z)$,

\be
W^{\alpha}\,W^{\beta}\,T_{\alpha,\beta}(z) \geq 0. 
\ee

\ni Acronym: WEC. Interpretation: the energy density as measured by any observer 
is everywhere positive.

{\bf The Dominant Energy Condition}: The WEC holds and $W_{\alpha}\,T^{\alpha,\beta}$ is everywhere
non-spacelike, equivalent to:

\be
T^{0,0} \geq |T^{\alpha,\beta}|\ph\ph\hbox{for $\alpha,\,\beta,\,= 1,2,3.$}
\ee

\ni Acronym: DEC. Used by \HE\ to show that matter cannot move faster than light (pp.s 91--94).

{\bf The Strong Energy Condition}: For all time-like $W_{\alpha}$,

\be
W^{\alpha}\,W^{\beta}\,T_{\alpha,\beta}(z) \geq \frac{1}{2}\,W^{\alpha}\,W_{\alpha}\,T(z),
\label{sec}
\ee

\ni ($T$ stands for the trace: $T^{\alpha}_{\alpha}$.) Acronym: SEC. Used to establish
the existence of BH singularities.

Turning to the NLQM, if we ignore terms involving 
$\partial\,h^{\mu,\nu}/\partial\,g^{\alpha,\beta}$, the \emt\ becomes:

\bar
\no T_{\alpha,\beta}(z) &\=& c\,\left\{
<\partial_{z,\alpha}\,\psi\,|
\,\partial_{z,\beta}\,\psi>^{(1)} \-
\half\,g_{\alpha,\beta}(z)\,g^{\mu,\nu}(z)\, 
<\partial_{z,\mu}\,\psi\,|
\,\partial_{z,\nu}\,\psi>^{(1)}\,\right\}\\ 
\no && \-\,g_{\alpha,\beta}(z)\,I(z).\\
&&
\ear

\ni where

\be
I(z) \= I_a(z) + I_b(z) + I_c(z).
\ee

Focusing on the SEC, let $W^{\alpha}$ be a unit time-like vector. 
From the terms other than $I(z)$
we find:

\bar
\no && W^{\alpha}\,W^{\beta}\,T_{\alpha,\beta}(z)\- \half\,W^{\alpha}\,W_{\alpha}\,T(z)
\=\\
\no && <\,|\,W^{\alpha}\,\btd_{\alpha,z}\,\psi\,|^2\,>^{(1)} 
\- \half\,W^{\alpha}\,W_{\alpha}\,<\,\btd_{\alpha,z}\,\psi\,\btd^{\alpha,z}\,\psi\,>^{(1)} \\
\no && \- \half\,W^{\alpha}\,W_{\alpha}\,\left\{\,
<\,\btd_{\alpha,z}\,\psi\,\btd^{\alpha,z}\,\psi\,>^{(1)}\right. \\
\no &&\left. \- \half\,4\,<\,\btd_{\alpha,z}\,\psi\,\btd^{\alpha,z}\,\psi\,>^{(1)}\,\right\} \\
\no && \= <\,|\,W^{\alpha}\,\btd_{\alpha,z}\,\psi\,|^2\,>^{(1)}\\ 
\no && \geq 0.\\
 &&
\ear

Thus, ignoring all the interesting terms, the SEC would hold. (A familiar result for the $N = 1$
massless scalar field, \HE\ p.95. The massive case can violate the SEC, although \HE\ argue 
only at extreme curvatures, which seems to beg the question as the SEC is vital to proving the
existence of such curvatures.)

The contribution to (\ref{sec}) from the $I$-terms is simply $\half\,I(z)$.
Let's first discuss $I_c(z)$.
We should not, however, examine this term for general wavefunctions, as 
the only relevant $\psi$ satisfy the
dynamical equations of section \ref{dynsect}. 
Suppose the `$a$' and `$b$' terms aren't in the Lagrangian. Integrating by parts on $x$ in (\ref{emtc2}) we can rewrite it as: 

\bar
\no I_c(z) &\=& 
 \- c\,\int\,dv(x)\,g^{\mu,\nu}(x)\,\left(\,\sum_{i=1}^{N-1}\,\prod_{j\neq i}\,
\int\,dv(x_j)\,\left[\,\psi^*\,\btd_{x,\mu}\,
\btd_{x,\nu}\,\psi\,\right]_{x_i = z}\, \right).\\
&&
\ear

\ni which vanishes by the dynamical equation, $D_c\,\psi = 0$. Hence, with 
no `$a$' or `$b$' terms, the SEC holds.

Adding the `$a$' term in the Lagrangian, 
integrating by parts in $I_a$ where a term contains a derivative integrated over,
reconstructs
either $D_a\,\psi$ or $D_a\,\psi^*$; hence $I_a(z) = 0$. If we have both `$a$' and `$c$' terms, 
note that $D_a\,\psi = 0$ implies $D_c\,\psi = 0$ at least in 
Minkowski space, so by the above arguments, the SEC would still hold. 
(This argument presumes that $D_a\psi + D_b\psi = 0$
has a unique solution.) Finally, consider $I_b(z)$. Again, 
IBPs can construct terms containing $D_b\,\psi^*$ or $D_b\,\psi$.
However, it does not follow from $D_a\,\psi = 0$, or from $D_a\psi+ D_b\psi+ D_c\psi = 0$,
that $D_b\,\psi = 0$. Morever, although these expressions may occur in the summed Lagrangian,
they appear with different integrations and prefactors, so don't align into a sum. 
Thus, $I(z)$ may make a contribution.

Finally, let's consider the terms involving the derivatives of the bivector with 
respect to the metric. Several possibilities arise. Adopting the construction of $h^{\mu,\nu}$
from the Appendix,
no such terms appear. In other cases they might contribute. 
 For one instance, adopt the parameter values
mentioned in section \ref{flatsect}. Then for the sum of leading terms (order $N$)
assemble into an expression:

\bar 
\no && \int\,\int\,dv_2(x,y)
\,\left\{\,<\partial_{x,\mu}\,\psi\, - <\,\partial_{x,\mu}\,\psi\,|\,\psi\,>^{(1)}\,|\,
\partial_{y,\mu}\,\psi\, - <\,\psi\,|\,\partial_{y,\mu}\,\psi\,>^{(1)}\,>^{(2)}\right\}\,\times\\
\no &&\,\frac{\delta\, h^{\mu,\nu}(x,y)}{\delta\,g^{\alpha,\beta}(z)};\\
&&
\ear 

So we are motivated to hypothesize about the matrix:

\be
M_{\alpha,\beta} \equiv 
\int\,\int\,dv_2(x,y)\,
\xi_{\mu}^*(x,y)\,\xi_{\nu}(x,y)\,\frac{\delta\, h^{\mu,\nu}(x,y)}{\delta\,g^{\alpha,\beta}(z)},
\ee

\ni for any (complex) bivector $\xi(x,y)$. We might postulate that $M$ is positive definite (in 
a certain sense): for any unit time-like vector $W^{\alpha}$, this contribution appears 
in the SEC condition:

\bar
\no && W^{\alpha}\,W^{\beta}\,T_{\alpha,\beta}(z)\- \half\,W^{\alpha}\,W_{\alpha}\,T(z) \=\\
\no &&  W^{\alpha}\,W^{\beta}\,M_{\alpha,\beta} + \half\,g^{\alpha,\beta}\,M_{\alpha,\beta}\\
\no && \geq 0.\\
&&
\ear

Of course, these terms might contribute with the opposite sign or nothing definite,
depending on the exact construction of the bitensor from the metric.

\section{Discussion\label{discsect}}

Space-time \wfs\ raise the spectre of a ``time cat", meaning a (macroscopic) object whose
\wf-computed dispersion is larger temporally than the object's natural duration.
Both ``time cats" and reverse causation (the future influencing the past) might be eliminated
by adopting a proposal from the ``multi-time \wf" literature (see \cite{lpt}): restricting
domains of the integrals to a subset of $M^N$, 
call it SLR (for ``space-like related"), of N-tuples such that
$x_i$ and $x_j$ are space-like related for all $i\,\neq\,j$. 
A complication of this proposal
stems from the fact that this subset has a boundary
 (made up of unions of light-cones).
For the Lagrangian program this means that in the derivation of the dynamical equations 
the integrations-by-parts that remove derivatives from
$\Delta\,\psi$ will generate boundary terms. These might be prevented by requiring ``allowable"
 $\psi$ to vanish on the boundary. 
But for the \emt\ another difficulty arises: SLR clearly depends
on the metric. So varying it will produce some complications. 

Where space-and-time cats are concerned, in NLQM this restriction step may be not necessary. 
Assume that, with suitable $a$, $b$, and $c$, $T^{\mu,\nu}$ reassembles into a term representing
dispersion of the energy-momentum. 
Energy bounds can then limit the extent of dispersion, whether in
spatial or temporal directions.

Concerning energy inequalities and space-time singularities: is it possible that,
in certain circumstances, the SEC might be violated? As a test case, let's put $N = 2$
and examine  

\bar
\no I_b(z) &\=&  b\,\left\{\, \int\,dv(y)\,h^{\mu,\nu}(z,y)
\left[\,<\,\partial_{x,\mu}\,\psi\,|\,\psi\,>^{(1)}\,\right]_{x=z} \,
<\,\psi\,|\,\partial_{y,\nu}\,\psi\,>^{(1)}\+\right.\\
\no && \left. \int\,dv(x)\,h^{\mu,\nu}(x,z)\,<\,\partial_{x,\mu}\,\psi\,|\,\psi\,>^{(1)}\,\left[
<\,\psi\,|\,\partial_{y,\nu}\,\psi\,>^{(1)}\,\right]_{y=z}\,\right\}\\
&&
\ear

Clearly, if the dynamical equations do not block it, $I_b(z)$ may take either sign,
depending on the comparison of derivatives at $z$ and other points. Thus, if along an
observer's worldline inside a Black Hole the wavefunction 
should become expanded in an appropriate sense 
the SEC might be overcome and the singularity
prevented. Presumably curvature would drive this process, but I cannot say exactly 
how this would happen.

\section{Appendix: Deriving the Bitensor\label{appsect}}

The intuitive derivation arises from imagining an embedding of \st\ into a higher-dimensional
flat (Euclidean, perhaps pseudo-Reimannian) space. Isometric (metric-preserving) embeddings 
for pseudo-Reimannian manifolds were 
described in the 1970s by Clarke, \cite{clarke}, and Greene, \cite{greene}, 
extending results of Nash from the 1960s for Reimannian manifolds. (For a recent review, 
see Gromov 2017, \cite{gromov}.)
Let $F:\,M\longrightarrow E$ denote this 
mapping. $F$ is said to be isometric if:

\be
\sum_{k}\,\partial_{x,\mu}\,F^{k}(x)\,\hbox{sign}_k\,\partial_{x,\nu}\,
F^{k}(x) \= g_{\mu,\nu}(x)
\ee

\ni Here $\hbox{sign}_k = \pm\,1$. 
The sum may extend to a large number of terms; for example, 
Clarke's construction required 87, and Greene's 342. 
Given this mapping, the \bt\ is then definable as:

\be
h_{\mu,\nu}(x,y) \= \sum_{k}\,\partial_{x,\mu}\,F^{k}(x)\,
\hbox{sign}_k\,\partial_{y,\nu}\,
F^{k}(y).
\ee

An issue arising from this construction is uniqueness: formulas for the embeddings vary by author.
All physical theories must avoid hypothesis flexibility 
in order to generate unique preditions.
And there is the matter of those embarrassingly-high dimensional spaces. 
I would not wish to engage
in debate over whether the world actually has 87 dimensions. 
Accordingly, I describe next a natural, intrinsic geometrical construction of the \bt\ 
for ``almost all" pairs $(x,y)$, which will suffice for our purposes.

I require some restrictions on the geometry of the manifold which should nevertheless 
allow the theory to be formulated on ``generic" \sts. The aspect I refer to is ``geodesic
connectivity". Let $EP$ (for exceptional pairs) denote the set of pairs $(x,y)$ in $M\times M$
for which there are either no geodesics connecting them or an infinite number. I will say
a manifold-plus-metric is ``allowable" if $EP$ has measure zero in $M\times M$ equipped 
with the product measure volume, $dv(x)\,dv(y)$. 
A familiar illustration here is the globe of the Earth (i.e., a two-sphere). Any pair
of points that are not antipodal possess exactly two connecting great-circle segments 
(the geodesics), while antipodal pairs have an infinite number. 
But the latter form a manifold of dimension
two in a four-dimensional manifold (of pairs). 
Hence the globe passes the test, as will any smooth deformations. 

For \sts: intuitively, pairs of events that cannot be connected by a geodesic curve 
(perhaps not by any curves) shouldn't exist (but they do in some cases; 
e.g., de Sitter space, see \HE, p. 126),
while those which are connected by infinitely-many (a subset of
 the so-called ``conjugate pairs")
should satisfy some equations and hence must form a set of zero measure. 
Because the theory in this paper is derived from a Lagrangian which integrates over $M\times M$, 
the allowability-condition imposes no burden.

Here's the recipe for the \bt. Let $M$ be an allowable \st\ and $(x,y)\,\notin\,EP$. 
Given a tangent vector $X$
at $x$ and $Y$ at $y$ and a geodesic $\gamma(s)$ connecting $x$ to $y$: $\gamma(0) = x$ and 
$\gamma(s) = y$ for some $s$, parallel transport $X$ along $\gamma$ to y and take the inner product
with $Y$ using the local metric $g_{\mu,\nu}(y)$. Then do it in reverse, transporting $Y$ to $x$
and comparing in the local metric $g_{\mu,\nu}(x)$. Perform these functions for all connecting
geodesics and take the average. For the formal definition, 
let $\Gamma(x,y)$ denote the set of geodesics connecting $x$ to $y$ and let
$n(x,y) = \#\Gamma(x,y)$. Let $P(\gamma;x,y)$ denote the propagator mapping 
the tangent space at $x$ to that at $y$ by parallel transport along the geodesic $\gamma$.
Then:

\bar
\no h_{\mu,\nu}(x,y)\,X^{\mu}\,Y^{\nu} &\=& \frac{1}{2\,n(x,y)}\,\sum_{\gamma \in \Gamma(x,y)}\,
\left\{\,g_{\mu,\nu}(y)\,Y^{\nu}\,P(\gamma;x,y)^{\mu}_{\alpha}\,X^{\alpha} \+\right.\\
\no && \left.g_{\mu,\nu}(x)\ X^{\mu}\,P(\gamma;y,x)^{\nu}_{\alpha}\,Y^{\alpha}\,\right\}.\\
&& \label{heqn}
\ear

If this formula is adopted, the integrand in formula (\ref{star}) will contain Dirac delta
functions. Writing the contravariant form without the vectors as:

\bar
\no h^{\mu,\nu}(x,y) &\=& \frac{1}{2\,n(x,y)}\,\sum_{\gamma \in \Gamma(x,y)}\,
\left\{\,g^{\mu,\alpha}(y)\,P(\gamma;x,y)^{\nu}_{\alpha} \+\right.\\
\no && \left. \hbox{terms with $x\,\leftrightarrow\,y$ and $\mu\,\leftrightarrow\,\nu$}\,\right\},\\
&& \label{heqn2}
\ear

\ni we find:

\bar
\no \frac{\partial\,h^{\mu,\nu}(x,y)}{\partial\,g^{\alpha,\beta}(z)} &\=& \sqrt{-g}^{-1}(z)\,
\frac{1}{2\,n(x,y)}\,
\sum_{\gamma \in \Gamma(x,y)}\,
\Big\{\,g^{\mu,\rho}(y)\,
\frac{\partial\,P(\gamma;x,y)^{\nu}_{\rho}}{\partial\,
g^{\alpha,\beta}(z)}\,\delta(\,z \in \gamma\,) \+\\
\no && \delta_{\alpha}^{\mu}\,P(\gamma;x,y)^{\nu}_{\beta}\,\delta(\,x-z\,) \+\\
\no &&  \hbox{terms with $x\,\leftrightarrow\,y$ and $\mu\,\leftrightarrow\,\nu$}\,\Big\},\\
&& 
\ear

 \ni Here, ``$\delta(\,z \in \gamma\,)$" indicates that $z$ is 
contained in the image of the geodesic $\gamma$ 
for a value of the parameter $s$ less that value that hits $y$, if $\gamma(0) = x$ (and
{\em vice versa}).  

I have written these expressions to make manifest the symmetries of $h^{\mu,\nu}$, but in fact the
dual parallel propagations (leading to the 1/2 in the above formula) aren't necessary, due to
the fact that parallel transport preserves inner products. To see this, transport a vector $X \in T_x$ to $T_y$ along a geodesic $\gamma$ to become, say, $X'$. Given another vector $Y \in T_y$,
transport both $X'$ and $Y$ back to $T_x$ along the same geodesic to be, say, $X''$ and $Y'$. 
Then

\be
g_{\mu,\nu}(x)\,X^{\mu''}\,Y^{\nu'} =      
g_{\mu,\nu}(y)\,X^{\mu'}\,Y^{\nu}.
\ee

But since we used the same geodesic, in fact $X = X''$, making the second transport in
the definition of $h^{\mu,\nu}$ redundant. (Of course, 
if we transported the vectors back with a different geodesic, curvature would come in.)     

Equations for the derivative of the propagator with respect to
the metric can be derived from the differential equation systems
defining geodesics and parallel transport of vectors, but I do not pursue these here.

\end{document}